\documentclass{article}
\usepackage[utf8]{inputenc}
\usepackage{url}
\usepackage{mathrsfs}
\usepackage{bm}
\usepackage{algorithmic}
\usepackage{mathrsfs}
\usepackage{color}
\usepackage{booktabs}
\usepackage[comma, numbers]{natbib}

\usepackage{graphicx}
\usepackage{float} 
\usepackage{subfigure}
\usepackage{caption}
\usepackage{graphicx}
\usepackage{amsmath}
\usepackage{float}
\usepackage{graphicx}
\usepackage{float}
\usepackage{mathrsfs}

\usepackage{amssymb}
\usepackage{bm}

\title{Robust mixture regression with Exponential Power distribution}
\author{CHEN XIAO
Hong Kong Baptist University}
\date{July 2020}

\begin{document}

\maketitle
\begin{abstract}
Assuming an exponential power distribution is one way to deal with outliers in regression and clustering, which can increase the robustness of the analysis. Gaussian distribution is a special case of an exponential distribution. And an exponential power distribution can be viewed as a scale mixture of normal distributions. Thus, model selection methods developed for the Gaussian mixture model can be easily extended for the exponential power mixture model.
Moreover,
Gaussian mixture models tend to select more components than exponential power mixture models in real-world cases, which means exponential power mixture models are easier to interpret.
In this paper,
We develop analyses for mixture regression models when the errors are assumed to follow an exponential power distribution.  It will be robust to outliers, and model selection for it is easy to implement.
\end{abstract}

\section{Introduction}
Finite mixtures of regression models have been broadly used to deal with various heterogeneous data. In most cases, we assume the errors follow normal distributions, and parameters are estimated by the maximum likelihood estimator (MLE).  However, this model will be very sensitive to outliers. 

 There are many methods to handle outliers in mixture of linear regression models. Trimmed likelihood estimator (TLE)(Neykov et al.,2007)  uses the top $N\alpha$ data with the largest log-likelihood.
The choice of $\alpha$ is very important for the TLE. It should not be too small or too large. But it's not easy to choose $\alpha$
 Yao(2014) proposed a robust mixture of linear regression models by assuming the errors follow the t-distribution. Song(2016) dealt with the same problem using Laplace distribution.

Assuming an exponential power distribution is one way to deal with outliers in regression and clustering. (Salazar E et.al., 2012; Liang F et.al. 2007;  Ferreira, M.A.,2014).  Gaussian distribution and Laplace dsitribution are special cases of exponential distribution. leptokurtic distributions have heavier tails than Gaussian distributions, which provide protection against outliers. And an exponential power distribution is a scale mixture of normal distributions. This makes it easier to understand.

Outliers are not the only problem. Choose the model selection is also a challenge for mixture model. Gaussian mixture models tend to select more components than exponential power mixture models in real world case, which means the 
interpretability is greatly reduced.[7] One possible explanation is that exponential power distribution is a scale mixture of Gaussian distributions.

 In this paper, robust mixture regression with Exponential power distribution and model selection for it are introduced.
\section{Exponential Power(EP) Distribution}
The density function of the Exponential Power(EP)  distribution $(p>0)$ where the mean is 0 is
\begin{equation}
    f_{p}(e ; 0, \eta)=\frac{p \eta^{\frac{1}{p}}}{2 \Gamma\left(\frac{1}{p}\right)} \exp \left\{-\eta|e|^{p}\right\},
\end{equation}

where $\Gamma(\cdot)$ is the Gamma function.  
When $0<p<2$, the distributions are heavy tailed, which means it will provide protection against outliers. When p=1, Equation(1) defines Laplace distribution. When p=2, Equation(1) defines Gaussian distribution.

\section{Robust regression with mixture of  exponential power distribution}
We assume that $(x_i,y_i)$ follows one of the following linear regression model with probability $\pi_k$,
\begin{equation}
    y_i=x'_i \beta_k+e_{ik}\qquad k=1,\ldots,K
\end{equation}

where $\sum_k^K \pi_k=1$.
$e_{ik}$ follows Exponential
Power (EP) distributions $f_{p_{k}}\left(e_{i k} ; 0, \eta_{k}\right)$

Assume $\bm{\Theta}=\left\{\bm{\pi},\bm{\eta}, \bm{\beta}\right\}$

The complete likelihood function is
$$\mathbb{P}(\mathbf{y}; \mathbf{x} , \bm{\Theta})=\prod_{i }^N \prod_{k=1}^{K}\left[\pi_{k} f_{p_{k}}\left(y_{i}-x_{i}^{\prime} \beta_{k} ; 0, \eta_{k}\right)\right]^{z_{ik}}$$

where$z_{ik}\in\{0,1\}$ is  an indicator variable. It implies that $y_i$ comes form $k$th linear regression model. And $\sum_{k=1}^{K} z_{ik}=1$.

And the complete log-likelihood function is
$$\log \mathbb{P}(\mathbf{y}; \mathbf{x} , \bm{\Theta})=\sum_{i}^N \sum_{k=1}^{K} z_{i k}\left[\log \pi_{k}+\log f_{p_{k}}\left(y_{i}-x_{i}^{\prime} \beta_{k} ; 0, \eta_{k}\right)\right]$$

In practice, the number
of components has to be estimated. So we use penalty function to select K automatically.(Huang,  T et.al., 2017)
$$l(\Theta)=\log \mathbb{P}(\mathbf{y}; \mathbf{x} , \Theta)-P(\pi ; \lambda)$$
where$P(\pi ; \lambda)=n \lambda \sum_{k=1}^{K} D_{k} \log \frac{\epsilon+\pi_{k}}{\epsilon}$

  $\epsilon$ is a very small positive number,
$\lambda$ is a positive tuning parameter , and $D_{k}$ is the number$
\text { of free parameters in the } k^{t h} \text { component. }$.

\section{Algorithm}
We propose a Generalized Expectation–Maximization algorithm to solve
the proposed model.

In the $\mathrm{E}$ step,  conditional expectation of $z_{i j k}$ given $e_{i j}$ is computed by the Bayes' rule:
\begin{equation}
\begin{array}{l}
\qquad \gamma_{i k}^{(t+1)}=\frac{\pi_{k}^{(t)} f_{p_{k}}\left(y_{i}-x_{i}^{\prime} \beta_{k} \mid 0, \eta_{k}^{(t)}\right)}{\sum_{l=1}^{K} \pi_{l}^{(t)} f_{p_{l}}\left(y_{i}-x_{i}^{\prime} \beta_{k} \mid 0, \eta_{l}^{(t)}\right)} 

\end{array}
\end{equation}

$\begin{aligned} Q\left(\bm{\Theta}, \bm{\Theta}^{(t)}\right)=& \sum_{i,  k} \gamma_{i  k}^{(t+1)}\left[\log f_{p_{k}}\left(y_{i}-x_{i}^{\prime} \beta_{k}\mid 0, \eta_{k}^{(t)}\right)+\log \pi_{k}\right] \\ &-n \lambda \sum_{k=1}^{K} D_{k} \log \frac{\epsilon+\pi_{k}}{\epsilon} \end{aligned}$

Then in the M step, $\bm{\Theta}$  are updated by maximizing $Q\left(\bm{\Theta}, \bm{\Theta}^{(t)}\right)$.

\begin{equation}
\pi_{k}^{(t+1)}=\max \left\{0, \frac{1}{1-\lambda K D_{k}}\left[\sum_{i} \gamma_{i k}^{(t+1)}-\lambda D_{k}\right]\right\}
\end{equation}

\begin{equation}
\eta_{k}^{(t+1)}=\frac{N_{k}}{p_{k} \sum_{i,} \gamma_{i k}^{(t+1)}\left|y_{i}-x_{i}^{\prime} \beta_{k}\right|^{p_{k}}}
\end{equation}
where $N_k=\sum_{i, } \gamma_{i  k}^{(t+1)}$.

On ignorning the terms not involving ${\beta}_k$, we have
\begin{equation}
    Q(\beta_k)=-\sum_{i=1}^N\sum_{k=1}^K \gamma_{ik}^{(t+1)}\eta_k^{(t+1)}\left|y_{i}-x_{i}^{\prime} \beta_{k}\right|^{p_{k}}
\end{equation}

Then we use MM algorithm to estimate $\beta_k$
\begin{equation*}
\begin{split}
      \{(y_i-x'_i \beta_k)^{\prime}(y_i-x'_i \beta_k)\}^{\frac{p_k}{2}}\leq\{(y_i-x'_i \beta_k^{(t)})^{\prime}(y_i-x'_i \beta_k^{(t)})\}^{\frac{p_k}{2}}\\+\frac{p_k}{2}\{(y_i-x'_i \beta_k^{(t)})^{\prime}(y_i-x'_i \beta_k^{(t)})\}^{\frac{p_k}{2}-1}\left((y_i-x'_i \beta_k)^{\prime}(y_i-x'_i \beta_k)-(y_i-x'_i \beta_k^{(t)})^{\prime}(y_i-x'_i \beta_k^{(t)})\right)
      \end{split}\end{equation*}

Thus
\begin{equation}
    \hat{\beta_k}=\arg {min} _{\beta_k}\sum_{i=1}^N\sum_{k=1}^K \gamma_{ik}^{(t+1)}\eta_k^{(t+1)}W_{ik}^{(t)}(y_i-x'_i \beta_k)^{\prime}(y_i-x'_i \beta_k)\label{e1},
\end{equation}
where $W_{ik}^{(t)}=\frac{p_k}{2}
\{(y_i-x'_i \beta_k^{(t)})^{\prime}(y_i-x'_i \beta_k^{(t)})\}^{\frac{p_k}{2}-1}$.

Equation (\ref{e1}) leads to the update
\begin{equation}
\beta_k^{(t+1)}=\left(\sum_{i=1}^N\sum_{k=1}^K \gamma_{ik}^{(t+1)}\eta_k^{(t+1)}W_{ik}^{(t)}\bm{x}_i\bm{x}^{\prime}_i\right)^{-1}\left(\sum_{i=1}^N\sum_{k=1}^K \gamma_{ik}^{(t+1)}\eta_k^{(t+1)}W_{ik}^{(t) 
}\bm{x}_iy_i\right).
\end{equation}

We can also use Newton-Raphson method to updata $\beta$.
\section{Simulation}
In this section, we demonstrate the effectiveness of the proposed method by simulation study.

Since TLE performed worse than the robust mixture regression model based on the t-distribution [4] and exponential power distribution can cover Laplace distribution, we only compare

1.the robust mixture regression model based on the t-distribution (MixT),

2.the robust mixture regression model based on the exponetional power distribution (MixEP).

To assess different methods, we record the mean squared errors (MSE) and the bias of the parameter estimates. In terms of the mixture switching issues, we use label permutation that minimizes the difference between predicted parameters and the true
parameter values.

Example $1 .$ Suppose the independently and identically distributed samples $\left\{\left(x_{1 i}, x_{2 i}, y_{i}\right), i=1, \ldots, n\right\}$ are sampled from the model
$$
Y=\left\{\begin{array}{ll}
0+X_{1}+\epsilon_{1}, & \text { if } Z=1 \\
0-X_{1}+\epsilon_{2}, & \text { if } Z=2
\end{array}\right.
$$
where $Z$ is a component indicator with $P(Z=1)=0.5, X_{1} \sim N(0,1), X_{2} \sim N(0,1),$ and both $\epsilon_{1}$ and $\epsilon_{2}$ follow the same distribution as $\epsilon .$ We consider the following four cases for the error density function of $\epsilon$ :(Weixin    Yao et.al., 2014)

Case I: $\epsilon \sim t_{1}$ , which means $t$ -distribution with degrees of freedom 1 ,

Case II: $\epsilon \sim 0.95 \mathrm{N}(0,1)+0.05 \mathrm{N}\left(0,5^{2}\right)-$,which is a contaminated normal mixture, 

Case $\mathrm{III}: \epsilon \sim N(0,1)$ with $5 \%$ of high leverage outliers being $X_{1}=2+\epsilon,$ and $Y=10+\epsilon$

Case IV: $Y=\left\{\begin{array}{l}0+X_{1}+\gamma_1+\epsilon_{1} \\ 0-X_{1}+\gamma_2+\epsilon_{2}\end{array}\right.$ ,$\begin{aligned}
\epsilon \sim N(0,1), P\left(\gamma_{1} \in(4,6)\right)=P\left(\gamma_{ 2} \in(4,6)\right)=0.1,\end{aligned} $
standard normal
distribution with 10\% outlier contamination.

\begin{figure}[H]
\centering
\subfigure[MixEP]{
\begin{minipage}[t]{0.4\linewidth}
\centering
\includegraphics[width=3in]{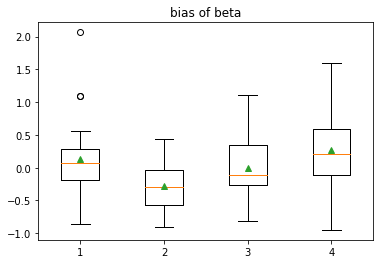}
\end{minipage}%
}%

\subfigure[MixT]{
\begin{minipage}[t]{0.4\linewidth}
\centering
\includegraphics[width=3in]{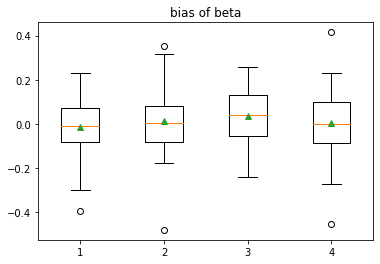}
\end{minipage}
}%
\centering
\caption{ Boxplots for the biass of the estimators of beta}
\label{fig:b4}
\end{figure}

CaseI:  The number of samples is N=400. 
\begin{figure}[H]
\centering
  \includegraphics[width=0.7\linewidth]{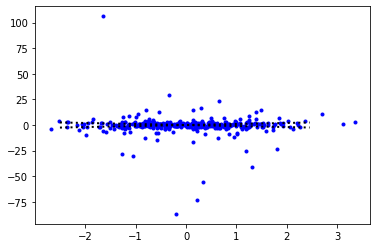}
  \caption{CaseI}
\label{fig:c5}
\end{figure}
Figure \ref{fig:c5} shows the simulation data in CaseI. We replicate this experiment 50 times, and the comparison of different methods are shown in Figure \ref{fig:b4} and Table \ref{t4}.

\begin{table}[H]
\centering
\begin{tabular}{lllll}
\hline
\multicolumn{5}{c}{MSE}                                                                                 \\ \hline
Method & \multicolumn{1}{c}{$\beta_{00}=0$} & \multicolumn{1}{c}{$\beta_{01}=1$} & \multicolumn{1}{c}{$\beta_{10}=0$} & \multicolumn{1}{c}{$\beta_{11}=-1$}  \\ \hline
MixT   & 0.0169& 0.01936& 0.01491& 0.0203\\ 
MixEP  & 0.3973& 0.2175& 0.2594& 0.4337            \\ \hline
\multicolumn{5}{c}{bias}                                                                                \\ \hline
Method & \multicolumn{1}{c}{$\beta_{00}=0$} & \multicolumn{1}{c}{$\beta_{01}=1$} & \multicolumn{1}{c}{$\beta_{10}=0$} & \multicolumn{1}{c}{$\beta_{11}=-1$} \\   \hline  
MixT   & -0.0169& 0.01393& 0.03333&0.001208       \\
MixEP  &
0.13677&-0.2754& -0.007769& 0.2658\\ \hline          
\end{tabular}
\caption{CaseI:The mean squared errors (MSE)and the bias of the parameter estimates for each estimation method.}
\label{t4}
\end{table}

\begin{figure}[H]
\centering
\subfigure[MixEP]{
\begin{minipage}[t]{0.4\linewidth}
\centering
\includegraphics[width=3in]{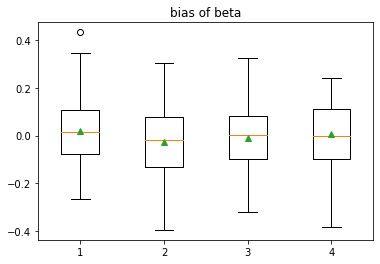}
\end{minipage}%
}%

\subfigure[MixT]{
\begin{minipage}[t]{0.4\linewidth}
\centering
\includegraphics[width=3in]{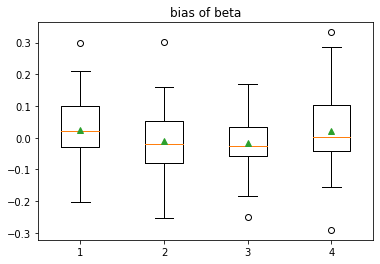}
\end{minipage}
}%
\centering
\caption{ Boxplots for the biass of the estimators of beta}
\label{fig:b2}
\end{figure}

CaseII:  The number of samples is N=400. 
\begin{figure}[H]
\centering
  \includegraphics[width=0.7\linewidth]{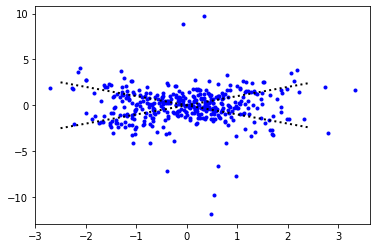}
  \caption{CaseII}
\label{fig:c1}
\end{figure}
Figure \ref{fig:c1} shows the simulation data in CaseII. We replicate this experiment 50 times, and the comparison of different methods are shown in Figure \ref{fig:b2} and Table \ref{t2}.

\begin{table}[H]
\centering
\begin{tabular}{lllll}
\hline
\multicolumn{5}{c}{MSE}                                                                                 \\ \hline
Method & \multicolumn{1}{c}{$\beta_{00}=0$} & \multicolumn{1}{c}{$\beta_{01}=1$} & \multicolumn{1}{c}{$\beta_{10}=0$} & \multicolumn{1}{c}{$\beta_{11}=-1$}  \\ \hline
MixT   & 0.0111& 0.0105& 0.006161& 0.01433        \\ 
MixEP  & 0.0189& 0.02219& 0.01943& 0.02024               \\ \hline
\multicolumn{5}{c}{bias}                                                                                \\ \hline
Method & \multicolumn{1}{c}{$\beta_{00}=0$} & \multicolumn{1}{c}{$\beta_{01}=1$} & \multicolumn{1}{c}{$\beta_{10}=0$} & \multicolumn{1}{c}{$\beta_{11}=-1$}\\   \hline  
MixT   & 0.02289& -0.009907& -0.01731& 0.02171           \\
MixEP    &   -0.0116& -0.0408& 0.02623& 0.03788\\\hline          
\end{tabular}
\caption{CaseII:The mean squared errors (MSE)and the bias of the parameter estimates for each estimation method.}
\label{t2}
\end{table}

\begin{figure}[H]
\centering
\subfigure[MixEP]{
\begin{minipage}[t]{0.4\linewidth}
\centering
\includegraphics[width=3in]{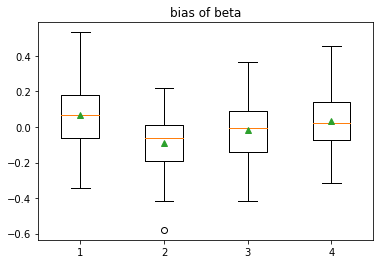}
\end{minipage}%
}%

\subfigure[MixT]{
\begin{minipage}[t]{0.4\linewidth}
\centering
\includegraphics[width=3in]{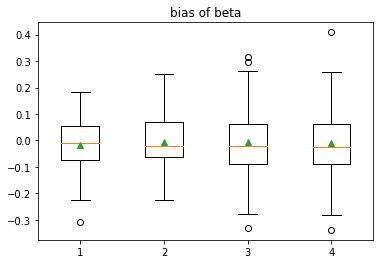}
\end{minipage}
}%
\centering
\caption{ Boxplots for the biass of the estimators of beta}
\label{fig:b3}
\end{figure}

CaseIII:  The number of samples is N=400. 
\begin{figure}[H]
\centering
  \includegraphics[width=0.7\linewidth]{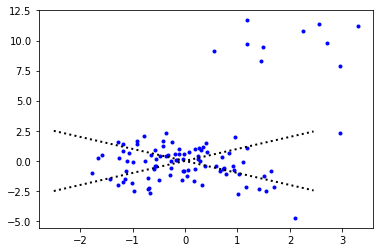}
  \caption{CaseIII}
\label{fig:c2}
\end{figure}
Figure \ref{fig:c2} shows the simulation data in CaseIII. We replicate this experiment 50 times, and the comparison of different methods are shown in Figure \ref{fig:b3} and Table \ref{t3}.

\begin{table}[H]
\centering
\begin{tabular}{lllll}
\hline
\multicolumn{5}{c}{MSE}                                                                                 \\ \hline
Method & \multicolumn{1}{c}{$\beta_{00}=0$} & \multicolumn{1}{c}{$\beta_{01}=1$} & \multicolumn{1}{c}{$\beta_{10}=0$} & \multicolumn{1}{c}{$\beta_{11}=-1$} \\ \hline
MixT   & 0.01188& 0.01052& 0.01857&0.02060\\ 
MixEP  & 0.03807& 0.03744& 0.02973& 0.02881            \\ \hline
\multicolumn{5}{c}{bias}                                                                                \\ \hline
Method & \multicolumn{1}{c}{$\beta_{00}=0$} & \multicolumn{1}{c}{$\beta_{01}=1$} & \multicolumn{1}{c}{$\beta_{10}=0$} & \multicolumn{1}{c}{$\beta_{11}=-1$}\\   \hline  
MixT   & -0.01858& -0.00432& -0.006022&-0.008520     \\
MixEP  & 0.06676& -0.08744& -0.01664& 0.03178\\ \hline          
\end{tabular}
\caption{CaseIII:The mean squared errors (MSE)and the bias of the parameter estimates for each estimation method.}
\label{t3}
\end{table}

\begin{figure}[H]
\centering
\subfigure[MixEP]{
\begin{minipage}[t]{0.4\linewidth}
\centering
\includegraphics[width=3in]{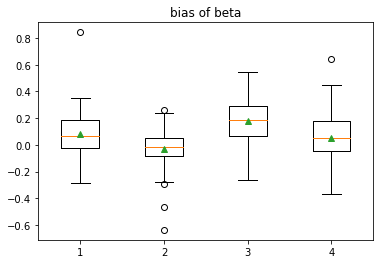}
\end{minipage}%
}%

\subfigure[MixT]{
\begin{minipage}[t]{0.4\linewidth}
\centering
\includegraphics[width=3in]{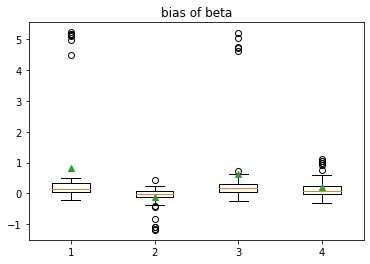}
\end{minipage}
}%
\centering
\caption{ Boxplots for the biass of the estimators of beta}
\label{fig:boat2}
\end{figure}

CaseIV:  The number of samples is N=400. 
\begin{figure}[H]
\centering
  \includegraphics[width=0.7\linewidth]{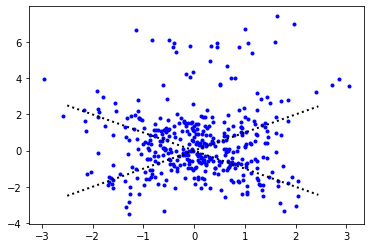}
  \caption{CaseIV}
\label{fig:boat1}
\end{figure}
Figure \ref{fig:boat1} shows the simulation data in CaseIV. We replicate this experiment 50 times, and the comparison of different methods are shown in Figure \ref{fig:boat2} and Table \ref{t1}.

\begin{table}[H]
\centering
\begin{tabular}{lllll}
\hline
\multicolumn{5}{c}{MSE}                                                                                 \\ \hline
Method & \multicolumn{1}{c}{$\beta_{00}=0$} & \multicolumn{1}{c}{$\beta_{01}=1$} & \multicolumn{1}{c}{$\beta_{10}=0$} & \multicolumn{1}{c}{$\beta_{11}=-1$} \\ \hline
MixT   &3.6119&0.1376&2.4279&0.1662     \\ 
MixEP  & 0.0411& 0.02742& 0.06486&0.04627         \\ \hline
\multicolumn{5}{c}{bias}                                                                                \\ \hline
Method & \multicolumn{1}{c}{$\beta_{00}=0$} & \multicolumn{1}{c}{$\beta_{01}=1$} & \multicolumn{1}{c}{$\beta_{10}=0$} & \multicolumn{1}{c}{$\beta_{11}=-1$}\\   \hline  
MixT   &0.8222&-0.1153& 0.6159&0.1979     \\
MixEP  & 0.08276& -0.02965& 0.1790& 0.05284    \\ \hline          
\end{tabular}
\caption{CaseIV:The mean squared errors (MSE)and the bias of the parameter estimates for each estimation method.}
\label{t1}
\end{table}

 It's hard to choose the dimension of freedom for Mixture of t-distribution. In the experiments shown above, the freedom leading to best results are used. However, the freedom selected by BIC could lead to bad results. In addition, selecting the number of components is also a problem for Mixture regression with t-distribution.

\section{Conclusion}
 The adaptively trimmed version of MIXT worked  well when  there are high leverage
outliers, but it can deal with drift.
We can't select a proper dimension of freedom for MIXT using BIC when faced with drift.

Adaptively removing abnormal points can improve the estimation accuracy in MixEP. It's a natural idea is to build a model to cover the drift. This will be the future work. 
\renewcommand{\bibname}{References}

\end{document}